\documentstyle[aps,manuscript]{revtex}
\begin{document}

\title{Attractive Boson and the Gas-Liquid Condensation}  

\draft
\author{Shun-ichiro Koh}
\address{ Physics Division, Faculty of Education, Kochi University  \\
        Akebono-cho, 2-5-1, Kochi, 780, Japan
        \thanks{e-mail address: koh@cc.kochi-u.ac.jp }
        }
\date{\today}

\maketitle
\draft

\maketitle
\begin{abstract}

  We calculate a grand partition function of the attractive Bose gas in the infinite 
space  within some approximations. Using the idea 
of the Yang-Lee  zeros, it is proved that 
the gas-liquid condensation occurs before the conventional condition of
the Bose-Einstein condensation is satisfied. 
 Further, it is pointed out that  Bosons with a zero 
momentum play a role of a trigger to this gas-liquid condensation. 
We discuss its implication to the trapped atomic gas. 
\end{abstract}


The study of the relationship between the Bose-Einstein condensation
(BEC) and the gas-liquid condensation (GLC) is a long-standing problem \cite{kah}.
 Recently the experimental realization of BEC  in dilute atomic vapors
changed this academic problem to a realistic one \cite{par}.
 Normally, the BEC, a  condensation into a lowest energy level 
 at low temperature and
high density, is thought to be an essentially different phenomenon
from the GLC in following points:
(1) The BEC is caused by the Bose statistics, 
not by the interparticle
interaction as in the GLC.
(2) The BEC is sometimes described as a condensation 
in momentum space, while the GLC occurs in coordinate space.

An interesting point is that 
the reason by which we distinguish them is not so obvious as it looks.
Both condensations have a similar  thermodynamic manifestation 
as a first-order phase transition. 
Further, in two types of the quantum gas, 
the attractive Fermi gas and the attractive Bose gas, 
the relationship between the BEC and the GLC manifests itself quite differently. 

For Fermions, the circumstance with low temperature and high density 
stabilizes 
the Cooper pairs: a BEC in a general sense \cite{coo}. Because of the 
Fermi statistics, however, two Fermions experience a strong repulsive force 
in the short distance.
Hence, when the attractive force is increased within 
the Bardeen-Cooper-Schriefer
(BCS) model, the GLC is impossible, a rigorous proof of which is
given recently \cite{koh}. 

For Bosons,  they remain in the gas state at high temperature 
and low density, because the quantum statistics plays a minor role.  
At low temperature, however, the Bosons have
no large positive kinetic energy  to stabilize
 the system behavior, especially in the BEC state. Thus, 
the attractive force plays a dominant role, so that the
compressibility is no longer positive definite. 
The dilute Bose system will 
collapse into the dense one, leading to the GLC.
Conversely, when the particle 
density increases by the GLC, an overlapping of the wave function is 
likely to cause the BEC.

There are some kinematical or thermodynamical evidences 
suggesting the instability 
of the attractive Bose gas.

(1) The Bogoliubov model says that if the interaction 
between the Bosons is attractive, the velocity
of sound propagating on the Bose-Einstein condensate 
would be imaginary, corresponding to
a divergence of the density fluctuation and leading 
to a drastic change of the whole system
\cite{bog}. 

(2)A variational argument of the many-Boson system shows that
 a many-body self-binding state (liquid) is more likely to
be stable than a gas of the bound dimers \cite{bru}. 

 (3) Recent stability analysis of the Gross-Pitaevskii 
equation in the trapped atomic gas with the attractive interaction
indicates a collapse of the BEC  at the thermal equilibrium.

 These features suggest that an another type of 
 relationship between the 
BEC and the GLC exists in the
attractive Bose gas. The GLC
in the classical imperfect gas is a long-standing and difficult problem 
 \cite{may}. 
In the attractive Bose gas,  however, there is 
a clear physical reason for the instability
 originated from the Bose statistics.
In view of this, at least for the instability mechanism to the liquid,
a model of the GLC in the attractive Bose gas would be
simpler than that in the imperfect classical gas or in the attractive Fermi gas.
With this in mind, it seems quite natural 
to consider the BEC and GLC on a common ground.

In this paper, we consider a spinless Bose gas
with a repulsive core represented by $H_{re}$ and a weak attractive
s-wave  pairing interaction $g (<0)$ \cite{pai}:
\begin{equation}
H=\sum_{k}\epsilon_{k}a^{\dagger}_{k}a_{k}+H_{re}+
\frac{g}{V}\sum_{k,k'}a^{\dagger}_{k}
a^{\dagger}_{-k}a_{-k'}a_{k'}. 
\label{}
\end{equation}
(In Eq.(1), $H_{re}$ is included only to assure universal natures of the 
many-body system, so that this paper does not deal with it explicitly.)
We assume that the $H_{re}$ and the attractive interaction 
 makes an additive contribution to 
the free energy because of  their different effective ranges, 
with a result that the grand partition function $Z(\mu)$ 
is factorized as $Z_0(\mu)Z_{re}(\mu)Z_{at}(\mu)$.
Further we assume that diluteness of the system allows a contact 
interaction in $g$ as a first approximation. 
What concerns us is a singularity induced by $Z_{at}(\mu)$.
We point out the following: (i) When the temperature
decreases and the density increases, 
the grand partition function of the attractive Bose gas becomes zero
at a negative critical value of the chemical potential
$\mu _c (<0)$ . 
(ii) The Bosons with a zero momentum play a special role in this instability.

 The GLC can be considered as a singularity
in the isothermal $p-V$ diagram. The pressure $p$ and the density
 $\rho$ are given by,
 \begin{equation}
Ê
\frac{p}{kT}=\lim_{V\to\infty}\frac{\ln Z_V}{V},
Ê
\end{equation}
\begin{equation}
\frac{\rho}{kT}=\lim_{V\to\infty}\frac{\partial}{\partial\mu}
ÊÊÊÊÊÊÊÊÊÊÊÊÊÊÊÊ \left(\frac{\ln Z_V}{V}\right),
\end{equation}
where $Z_{V}$ is the grand partition function in the volume $V$
 and $\mu$ is a chemical potential. 
A zero of $Z(\mu)$ defines the GLC,  giving the isothermal line a continuous 
but not differentiable point (Yang-Lee's zero \cite{yan}).
On the contrary, a pole of $Z(\mu)$ defines the BEC. 
Because of the
logarithmic function in Eqs.(2) and (3), the pole gives the isotherm near 
the BEC point a similar shape
 to that by the zero. In contrast to
the GLC, however, the isotherm is not only  continuous 
but also differentiable at the BEC point of the free Bose gas. 
 
The $Z(\mu)$ which explains the GLC
should have a following feature. Generally in the condensed matter, 
local changes
induced by an external perturbation, or local interactions responsible 
for the thermodynamic quantities in the normal phase are well described 
by a sum of the disconnected ring diagram like Fig.1(a). 
To describe a macroscopic change like the GLC, 
however, it is 
essential to include much greater networks of the interaction 
extending to all particles which consist of the system.
 As the particle number increases, 
a variety of such globally connected diagrams increases  rapidly.
Further, this connected networks is necessary for satisfying 
the quantum statistics correctly.
When two particles having a same momentum and 
belonging to different ring diagrams are exchanged, 
a connected graph is produced as a result (for Fig.1(a), 1(b) is produced). 
Hence, to include a symmetrized or antisymmetrized state into calculation, 
the sum of the disconnected ring diagrams 
must be supplemented by the
corresponding connected diagram. 
Summing up all these diagrams for $Z(\mu)$ is an
essential step to describe the macroscopic change caused by
the quantum statistics such as  the GLC in the 
quantum system. (This viewpoint was originally formulated in the case of the 
attractive Fermi gas: superconductivity  \cite{gau} \cite{lan}.)

For completeness, we repeat the formalism in Ref.\cite{gau} \cite{lan}
in a simplified form, and apply it to the Boson system.
The connected diagram like Fig.1(b) is made of a combination of elementary diagrams
like Fig.1(c) consisting of 2$s$ particle lines.
Since each graph in Fig.1(c) consists of Boson lines with a single $(l,p)$, 
 a sum over  $(l,p)$,
\begin{equation}
	 K_s=\frac{1}{V¥}\sum_{l,p}
	      \left(-\frac{1}{V¥}\frac{g}{\beta¥}¥¥
	             \frac{1}{(\epsilon_p-\mu)^2+(\frac{\pi l}{\beta¥})^2¥¥¥}¥\right)^s¥¥¥,
	\label{¥}
\end{equation}¥
is an elementary unit \cite{sig}.
(Since each interaction line is connected to two $K_s$, each $K_s$ has $(g/V)^s$.)
Only after summing the $K_s$'s over all ways of connecting
them  by the interaction lines, we get an expected $Z(\mu)$ .

Consider a connected diagram like Fig.1(b) where the $K_s$ appears $\nu_s$ times 
 as in $\{  \nu _s \} =(\nu_1,\nu_2, \ldots )$
 (In Fig.1(b),  $\{  \nu _s \} =(3,3,0, \ldots )$),
and the  $K_s$'s are connected to each other by $n$ interaction lines.

(1) There are $\nu _s!$ ways of rearrangement which leaves the diagram invariant.
(2) Since there are a number of ways of distributing  frequency $l$ and 
momentum $p$ to each  $K_s$,  
each interaction line connecting the  $K_s$'s has an individuality 
characterized by $(l,p)$ and $(l',p')$ of 
particles which enter or emerge at both ends. This allows us  $n!$ ways of rearrangements 
which produce different diagrams.

With this in mind, we get,
\begin{equation}
     \frac{Z(\mu)}{Z_0Z_{re}¥}=\sum_{\{\nu _s \}}n!\prod_{s}^{\infty¥}¥\frac{1}{\nu _s!¥}
                     \left(\frac{-K_s}{2s¥}¥\right)^{\nu _s}¥¥¥,	
	\label{¥}
\end{equation}¥
where $2s$ in the denominator is a number of rotations which leave the $K_s$ invariant.
If the sum over ${\nu _s}$ can be carried out independently to $n$, we 
simply obtain $\frac{Z(\mu)}{Z_0Z_{re}¥}=n!\prod_{s}^{\infty¥}exp(-\frac{K_s}{2s¥})¥¥¥$, but
in reality they are related to each other by $n=\sum_{s}s\nu _s¥$.
To include this constraint in the summation, and to transform $n!$  to a 
more tractable form, an identity
 \begin{equation}
 	n!=V\int_{0}^{¥\infty}dt(Vt)^ne^{-Vt}¥,
 	\label{¥}
 \end{equation}¥ 
is used, and $n$ in $(Vt)^n$ is replaced by $\sum_{s}s\nu 
 _s$. Using Eqs.(4) and (6) in Eq.(5), we can combine $K_s^{\nu _s}$ with $(Vt)^n$.
 Summing it over $\nu _s$, we get,
 \begin{equation}
 	 \frac{Z(\mu)}{Z_0Z_{re}¥}=V\int_{0}^{¥\infty}dt \exp(-Vt-\sum_{s}¥\frac{1}{2s¥}K_s'(t)),
 	\label{¥}
 \end{equation}¥
where $K_s'(t)$ is a redefinition of the $K_s$ by replacing $1/V$ by $t$ in the 
right-hand side of Eq.(4). Summing  $K_s'(t)/s$ over an integer $s$ from $1$ 
to $\infty$ using an identity,
\begin{equation}
     \ln(1+x)=-\sum_{m=1}^{\infty}\frac{(-x)^m}{m¥} ,	
	\label{¥}
\end{equation}¥
 and carrying out the summation in $K_s'(t)$ 
over an even integer $l$ including zero (Bose 
statistics) by the use of an identity, 
\begin{equation}
	\prod_{m=1}^{\infty}\left(1+\frac{z^2}{(2m)^2¥}¥\right)
	               =\frac{2}{\pi z¥}\sinh\frac{\pi z¥}{2¥},¥¥¥
	\label{¥}
\end{equation}¥
we obtain a final form,
\begin{eqnarray}
Z(&\mu&) = Z_0Z_{re}\int_{0}^{\infty} dt \exp(-Vt)\prod_{p=0}
               \left(1+\frac{gt}{\beta}\frac{1}{(\epsilon_p-\mu)^2}¥¥\right)
            \nonumber \\ 
	   & \times &    
            \left(\frac{\sinh \beta
                       \sqrt{(\epsilon_p-\mu)^2+\frac{gt}{\beta}¥}}
                  {\sinh \beta(\epsilon_p-\mu) }
                  \frac{(\epsilon_p-\mu)}{\sqrt{(\epsilon_p-\mu)^2+\frac{gt}{\beta}¥}}\right)^2
                  .¥¥¥¥	
	\label{¥}
\end{eqnarray}¥

In view of Eq.(10), the interaction does not produce a new pole in $Z(\mu)$, 
the conventional definition of the BEC. Instead, we obtain a new insight 
on the zero of $Z(\mu)$, the definition of the GLC.

To extract qualitative features of   $Z(\mu)$ from Eq.(10), 
a most important part of the integrand is 
$\left(1+\frac{gt}{\beta}\frac{1}{(\epsilon_p-\mu)^2}¥¥\right)$, which comes 
from $l=0$, and, because of $(\epsilon_p-\mu)^2>\mu^2$,
 especially important is its least term 
$\left(1+\frac{gt}{\beta}\frac{1}{\mu^2}¥¥\right)$ coming from a zero momentum.
When the negative $\mu$ increases, this term approaches zero first among 
many terms in the product because of $g<0$. 
Hence, at a critical value $\mu _c (<0)$ , the zero of $Z(\mu)$ is realized by a cancellation of two 
integrals which are obtained by dividing the integrand into two parts 
at this term. Generally the 
density $\rho$ derived by Eq.(3) is a monotonic increasing function 
of $\mu$, to which function the zero of this  
$Z(\mu)$ adds a discontinuity at $\mu =\mu _c$. Equations (2) and (3) 
together give the isotherm exhibiting the GLC.

It must be noted that the Bosons with the zero momentum play a role of a 
trigger to the GLC. The negative divergence of its $\rho$ in Eq.(3) 
implies that {\it the  Bosons  with the zero momentum
escape from the dilute gas, thereby making a droplet \/}.
This high density assembly of the zero momentum Bosons is in favor of the BEC.

Let us estimate a condition of this GLC for a  weak attractive force.
 As a first approximation, we 
can estimate $\mu _c$ by,
\begin{equation}
	\int_{0}^{\infty}dt \exp(-Vt)
	     \left(1+\frac{gt}{\beta}\frac{1}{\mu^2}¥¥\right)=0,
	\label{¥}
\end{equation}¥
and get a condition:$\mu _c\cong-\sqrt {|g|k_BT/V}$.
Further, for $\mu (T)$, we use the formula in the free Bose gas:
$\mu (T)=-(g_{3/2}(1)/2\sqrt{\pi})^2k_BT_{BEC}[(T/T_{BEC})^{3/2}-1]^2¥$, where
$g_{3/2}(x)=\sum_{m}x^m/m^{3/2}¥$.
Solving $-\sqrt {|g|k_BT/V}\cong \mu (T)$, we get,
   \begin{equation}
   	\mu _c\cong-\left(\frac{2\sqrt{\pi}k_BT_{BEC}}{g_{3/2}(1)¥}¥\right)^{2/5}
   	             \left(\frac{|g|}{V¥}¥\right)^{3/5} ,¥¥
   	\label{¥}
   \end{equation}¥
which shows non existence of the 
threshold strength of the attractive force:
{\it Under an infinite small attractive force, the GLC occurs before the 
chemical potential reaches zero \/}.  

For a  weak attractive force, we extend $n\lambda^3=g_{3/2}(1)$, 
the well known condition of the BEC, to a general condition of the instability
 caused by   $Z(\mu)$.
  ($n$ is a number density and $\lambda=(2\pi mk_BT/h^2)^{-1/2}$, a thermal wave length
  \cite{hua}.)
  Thus we  replace $\mu=0$ by  $\mu_c$ such that  
 $n\lambda^3=g_{3/2}(e^{\beta\mu _c})$.
 Because of $e^{\beta\mu _c}<1$, $T_{GLC}>T_{BEC}$ for a constant $n$, 
 and $n_{GLC}<n_{BEC}$ for a constant $T$.
  Figure.2 illustrates the condition 
 of the GLC by a solid curve for $g/V=-5nK$ and by a dotted curve for  
 $g/V=-5\mu K$ with $T(nK)=30\lambda^{-2}(\mu m)$ in Rb atom. 
 (The shaded area represents the BEC 
 phase of the free Bose gas: $n\lambda^3 \geq g_{3/2}(1)$.) 
 {\it When the temperature decreases and the density increases, the GLC occurs 
 before the conventional BEC condition  is satisfied \/}. Under the same 
 strength of the attractive force, it manifests itself more evidently in 
 the lower temperature region. As the strength increases, its position 
 in the phase diagram moves to the higher temperature and lower density region,
  a more realizable environment. Note that the GLC is very sensitive to 
  $g/V$ due to its cooperative nature.

The trapped atomic gas is the Boson system in the harmonic 
potential. The BEC in the confined space has been studied theoretically 
from old days. Although some differences from that in the infinite space 
were reported, they do not change the essential features of the BEC. 
Hence, for the trapped Bosons as well, we 
can expect an essentially similar phase diagram \cite{qua}.

The BEC with the attractive interaction was first observed in  $^7$Li gas 
as a metastable state \cite{bra}. 
A peculiar feature of the atomic gas is that we can control the sign of 
the interaction between the atoms using so-called ``Feshbach resonance''. 
Using this technique, the condensate over a wide range of 
interaction strength is realized . By a sudden switching of the 
interaction from the repulsive to the attractive one, a collapse of the 
BEC gas to a high density spot was predicted. Recently such a phenomenon 
was really observed in $^{85}$Rb \cite{cor}. When such an experiment is performed in 
the region of $n\lambda^3 < g_{3/2}(1)$ (the normal phase of the free Bose gas) 
with the number of the atoms  large 
enough to reach the thermal equilibrium, this type of experiment 
will provide us a method for finding the GLC line predicted in Fig.2.

In the attractive Bose gas, a possibility of the Boson pair has been 
explored by many people
in analogy with the Cooper pair \cite{noz}. In view of the result 
of this paper, however, the attractive force responsible for such a Boson 
pair must have a property such that there is no residual attractive 
interaction between different Boson pairs. When such a residual force is 
there, the GLC must occur instead of the Boson pair formation.
Forbidding the residual force, however, should impose a 
somewhat artificial constraint on the attractive interaction.
 
This paper deals with only an initial stage of the instability.
Once the Bose gas changes to the liquid, 
 the short range repulsive force due to $H_{re}$ plays a dominant role. 
 (When the Bosons are in a high density gas state at high temperature, 
 the situation
 is similar.) This makes Fig.2 a more complex phase diagram:
The critical point $(n_c, \lambda_c)$ will appear on the GLC line, 
and the line below the 
$(n_c, \lambda_c)$  must be replaced by the gas-liquid coexisting region.
 Inclusion of the repulsive force will reveal a more
complex nature of the later stage of this instability.
To understand whether the formation of the droplet immediately leads to 
the BEC or not, we must get the $Z(\mu)$ which includes the $H_{re}$ explicitly.
  The later stage of this instability is an open problem.

Acknowledgment
Ê
The author thanks K.Burnett and his group, K.Miyake and Y.Tsue 
for valuable discussion.

Ê


¥


%
%
 \begin{figure}
 \caption{ (a) A sum of the ring diagram which represents local changes.
       (b) A corresponding connected diagram which represents macroscopic changes.
       (c) Examples of building block  $K_s (s=1,2,3,4)$ of the connected 
       diagram, which are classified by $s$ where $2s$ is a number of particle lines.
       }
 \label{Fig.1}
 \end{figure}

 \begin{figure}
 \caption{$(n,\lambda)$ phase diagram of the attractive Bosons, 
 where $n$ is a  number density and 
$\lambda=(2\pi mk_BT/h^2)^{-1/2}$. A solid curve is a GLC line defined by 
$n\lambda^3=g_{3/2}(e^{\beta\mu _c})¥$ for $g/V=-5nK$, and a dotted curve for  
 $g/V=-5\mu K$ with $T(nK)=30\lambda^{-2}(\mu m)$ in Rb atom. 
 Shaded area is the BEC 
phase of the free Bose gas defined by $n\lambda^3 \geq g_{3/2}(1)$.
       }
 \label{Fig.2}
 \end{figure}
%
%


\begin{thebibliography}{99}

    \bibitem{kah} B.Kahn and G.E.Uhlenbeck, Physica,{\bf 5\/}, 399(1938).
       
    \bibitem{par} As a review, A.S.Parkins and D.F.Walls, Phy.Rep,
    {\bf  303\/}, 1 (1998), 
     F.Dalfovo, S.Giorgini, L.P.Pitaevskii and S.Stringari, Rev.Mod,{\bf 71}, 463(1999)
     
     \bibitem{coo} L.Cooper, Phy.Rev,{\bf 104\/},1189(1958)
Ê
     \bibitem{koh} S.Koh, Phys.Lett.A, {\bf 229\/},59(1997)
     
     \bibitem{bog} N.N.Bogoliubov, J.Phys.USSR, {\bf 11\/},23(1947)
     
      \bibitem{bru} L.W.Brush, Phys.Rev.B, {\bf 13\/},2873(1976)
      
      \bibitem{may} J.E.Mayer and M.G.Mayer, {\sl Statistical Mechanics\/}
      (John Wiely and Sons,New York, 1946)
      
       \bibitem{pai}Pairing interaction is a well approximation of the 
       general attractive interaction, and besides it simplifies later calculations.
       
      \bibitem{yan} C.N.Yang and T.D.Lee, Phys.Rev. B,{\bf 87}, 404(1952), and
       T.D.Lee and C.N.Yang, {\sl ibid\/}, 87, 410(1952)
       
      
      \bibitem{gau} M.Gaudin, Nucl.Phys.A,{\bf 20\/},513(1960)
      
      \bibitem{lan} J.S.Langer, Phys.Rev.A, {\bf 134\/},553(1964)
      
      \bibitem{sig}In contrast with Ref.[10][11], a minus 
      sign due to the Fermi statistics does not appear in Eq.(4). But 
      because of a different sign definition of $g$ in Eq.(1), Eq.(4) has 
      a similar form as that in  Ref.[10][11].
      
      \bibitem{hua}For example, K.Huang,Ê {\sl Statistical Mechanics\/},
       (Wiley, New York, 1987)
       
      \bibitem{qua} The quantum fluctuation caused by the confinement 
        within a bottom of the harmonic potential acts as a repulsive 
        interaction to Bosons, which may give rise to a complex nature 
        in the  attractive force  between the Bosons.
        
        
      \bibitem{bra} C.C.Bradley,  C.A.Sackett,  J.J.Tollett  and  R.G.Hulet, 
       Phys.Rev.Lett, {\bf 75\/},1687(1995)
     
          
       \bibitem{cor} S.L.Cornish, N.R.Claussen, J.L.Roberts,  E.A.Cornell 
       and  C.E.Wieman,  cond-mat/0004290
       
       
       
       \bibitem{noz} P.Noziere and D.Saint James, J.Physique,{\bf 43},
        1133(1982), and references therein.
        
       

\end{thebibliography}
\end{document}